\documentclass[twocolumn,prb,aps,floatfix]{revtex4}

\usepackage{graphicx}
\usepackage{dcolumn}
\usepackage{bm}

\begin{document}

\preprint{}

\title{Carrier relaxation mechanisms in self-assembled (In,Ga)As/GaAs quantum
  dots: Efficient $P\rightarrow S$ Auger relaxation of electrons}

\author{Gustavo A. Narvaez}
\affiliation{National Renewable Energy Laboratory, Golden, Colorado 80401}
\author{Gabriel Bester}
\affiliation{National Renewable Energy Laboratory, Golden, Colorado 80401}
\author{Alex Zunger}
\affiliation{National Renewable Energy Laboratory, Golden, Colorado 80401}

\date{\today}

\begin{abstract}
  We calculate the $P$-shell--to-$S$-shell decay lifetime $\tau(P\rightarrow
  S)$ of electrons in lens-shaped self-assembled (In,Ga)As/GaAs dots due to
  Auger electron-hole scattering within an atomistic pseudopotential-based
  approach. We find that this relaxation mechanism leads to fast decay of
  $\tau(P\rightarrow S)\sim 1-7\;{\rm ps}$ for dots of different sizes. Our
  calculated Auger-type $P$-shell--to-$S$-shell decay lifetimes
  $\tau(P\rightarrow S)$ compare well with data in (In,Ga)As/GaAs dots,
  showing that as long as both electrons and holes are present there is no
  need for an alternative polaron mechanism.
\end{abstract}

\maketitle

\section{Introduction}

Upon photoexcitation of an electron and hole in the barrier of an
(In,Ga)As/GaAs self-assembled quantum dot the carriers relax to their ground
states through a complicated dynamics. Much debate has taken place on the
mechanisms responsible for the final stages of the non-radiative decay
dynamics, which have been observed to involve relaxations of about
$40$-$60\;{\rm meV}$ and take place surprisingly fast---within $2$-$60\;{\rm
  ps}$. These decay times are much smaller that the radiative recombination
times $\tau_R\sim 1\;{\rm ns}$ observed in (In,Ga)As/GaAs
dots.\cite{karachinsky_APL_2004,buckle_JAP_1999,bardot_PRB_2005} To explain
this fast relaxation, three alternative mechanisms have been proposed and
supported by model calculations: multiphonon-emission, Auger carrier-carrier
scattering, and polaron decay. 
To provide a general perspective we first outline in this paper the general
decay channels of photoexcited carriers in the GaAs-barrier of (In,Ga)As/GaAs
self-assembled quantum dots (Sec. \ref{literature_review}), and then we focus
on the $P\rightarrow S$ Auger cooling due to electron-hole scattering,
providing accurately calculated results. We use a realistic atomistic,
pseudopotential-based approach (Sec. \ref{Auger_calculation}) that has been
recently applied to successfully reproduce the magnitude of the radiative
recombination lifetime of ground-state electrons and holes in (In,Ga)As/GaAs
dots (Ref. \onlinecite{narvaez_PRB_2005b}) and CdSe colloidal dots (Ref.
\onlinecite{califano_NanoL_2005}).
Our results for inter-shell decay time $\tau(P\rightarrow S)$ compare well
with data from experiments in which photoexcited holes are present. Thus, as
long as both an electron and hole are present the Auger mechanism can explain
fast inter-shell relaxation without resorting to other (e.g. polaronic decay
or multi-phonon emission) mechanisms.

\section{Characteristic dynamical processes of excited electrons and holes
  in self-assembled (In,Ga)As/GaAs quantum dots}
\label{literature_review}

One distinguishes first between systems having a lone carrier, either electron
or hole, and systems having both an electron and hole. A lone carrier can be
produced by doping the
dot\cite{bras_APL_2002,kammenerer_APL_2005,zibik_PhysicaE_2005,zibik_PRB_2004,zibik_PhysicaE_2005a,sauvage_PRL_2002}
or by electrochemical injection.\cite{guyot-sionnest_JCP_2005} Exciting a lone
carrier and following its decay
\cite{zibik_PRB_2004,zibik_PhysicaE_2005a,sauvage_PRL_2002} is a specialized
field and will be reviewed briefly in Sec. \ref{inter-shell_decay}. More
commonly we encounter relaxation of systems having both photoexcited electrons
and holes. This is reviewed next.
Figure \ref{Fig_1} sketches four non-radiative relaxation processes that take
place following photocreation of an electron-hole pair in an (In,Ga)As/GaAs
quantum dot system. The electron is shown as a solid dot and the hole as a
circle. The processes are illustrated with a dot with sparse confined electron
(CB) states $\{e_0,\,e_1,\,e_2\}$, and with a much denser set of confined hole
(VB) states $\{h_0,\,h_1,\,\ldots,\,h_k,\,\ldots,\,h_N\}$ as is characteristic
of self-assembled dots. The continuum of states of the wetting layer (dashed
region) and GaAs barrier (shaded) are also shown schematically. The main
observed carrier relaxation processes are the following.

\subsection{Barrier-to-wetting layer carrier capture}

Non-resonant photoexcitation of an electron-hole pair in the barrier [Fig.
\ref{Fig_1}(a)] often leads to capture by wetting-layer (WL) quasi-continua.
This process consists of carrier thermalization within the GaAs barrier and
subsequent capture by the WL. Barrier thermalization occurs within $1\;{\rm
  ps}$.\cite{siegert_PRB_2005,yuan_PhysicaB_1999} Siegert {\em et al.}
measured time-resolved photoluminescence (PL) signal from the wetting layer of
InAs/GaAs dots at high excitation and found a capture time of $\sim 2\;{\rm
  ps}$ regardless of doping (Ref.  \onlinecite{siegert_PRB_2005}).  Similarly,
in undoped dots, Sun {\em et al.} have found a capture time smaller that
$2\;{\rm ps}$ (Ref.  \onlinecite{sun_Nanotechnology_2005}), while Yuan {\em et
  al} observed a capture time of about $10\;{\rm ps}$ (Ref.
\onlinecite{yuan_PhysicaB_1999}).

\subsection{Carrier capture from the wetting layer into the dot}

\begin{figure}
  \includegraphics[width=8.5cm]{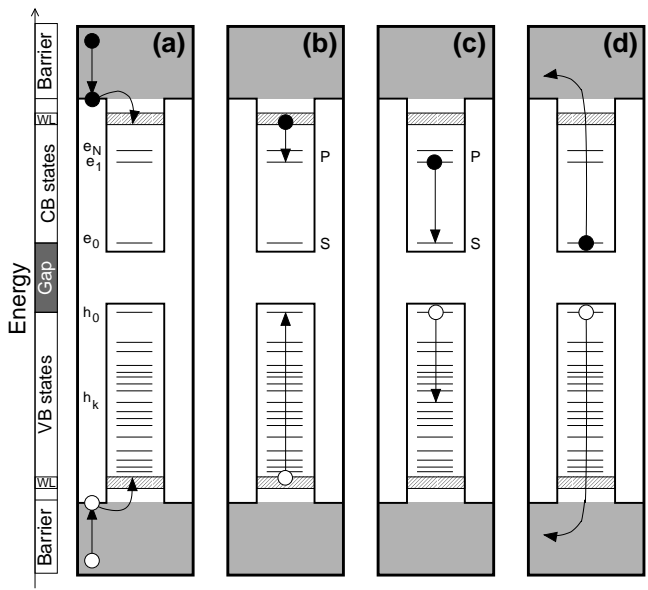}
\caption{{\label{Fig_1}}Sketch of different dynamical process experienced by
  photocreated carriers in a self-assembled (In,Ga)As/GaAs quantum dot: (a)
  Barrier-to-wetting layer (WL) carrier capture, (b) carrier capture from the
  wetting layer into the dot, (c) carrier relaxation within the dot, (d)
  thermal escape of carriers.}
\end{figure}

Following barrier-to-wetting layer carrier capture, the hole
relaxes to the lowest-energy confined hole state $h_N$ while the electron is
captured from the bottom of the wetting layer to the highest-energy confined
state [illustrated by $P$; Fig.  \ref{Fig_1}(b)].
Sosnowski {\em et al.}\cite{sosnowski_PRB_1998} found in time-resolved
differential transmission experiments at low excitation in an (In,Ga)As/GaAs
dot with two confined electron states that the electron capture time is
$2.8\;{\rm ps}$.
On the other hand, a {\em combined} capture time has been derived from
time-resolved photoluminescence (PL) experiments at high excitation by several
groups. (These times are affected by the subsequent intra-dot carrier
relaxation.) Siegert {\em et al.}\cite{siegert_PRB_2005} have found a capture
time of $4.9\;{\rm ps}$ in undoped dots, and $5.4\;{\rm ps}$ and $6.1\;{\rm
  ps}$ in {\em n}-doped and {\em p}-doped dots, respectively.\cite{note_00}
Similarly, Yuan {\em et al.} \cite{yuan_PhysicaB_1999} found a capture time
within $5\;{\rm ps}$, while Sun {\em et al.} found a capture time of less than
$2\;{\rm ps}$ (Ref. \onlinecite{sun_Nanotechnology_2005}).

%
%

\subsection{Relaxation of excited carriers within the dot}
\label{inter-shell_decay}

Following carrier capture from the wetting layer into the dot, carriers can
experience different dynamical processes.
These processes largely reflect the type of spacings that exist between
various confined states. The (In,Ga)As/GaAs system has interesting properties
in this respect. First, not only are these direct gap materials, but the
competing band-structure valleys ($X,L$) are rather far energetically from
$\Gamma$ [unlike InP or PbSe (Ref. \onlinecite{landolt_bornstein_table})], so
these materials, specially InAs, are in fact {\em strongly} direct-gap
systems.
Second, the hole mass in InAs is much heavier than the electron mass, so
confined hole states tend to be more densely spaced than electron states.
Third, the electron states are arranged in $S,\;P,\;D\, \dots$ ``shells'' and
each shell shows intra-shell level splittings, e.g. ${\cal E}(P_1)\neq {\cal
  E}(P_2)$ are split by $1$-$6\;{\rm meV}$, while inter-shell splittings are
larger, e.g. $S$-$P$ spacing is $40$-$60\;{\rm meV}$ (Refs.
\onlinecite{bras_APL_2002,kammenerer_APL_2005,zibik_PRB_2004,sauvage_PRL_2002,zibik_PhysicaE_2005,muller_APL_2003})
(compared to $\sim 300\;{\rm meV}$ in CdSe dots). Thus, the intra-shell
splitting is of the order of (small wave vector) acoustic phonon energies,
whereas inter-shell spacing is larger than (small wave vector) longitudinal
optical phonon energies.
Therefore, inter-shell relaxation via single-phonon emission due to
electron-phonon coupling (within the Born-Oppenheimer adiabatic approximation)
is expected to be ineffective---\cite{bockelmann_PRB_1990,benisty_PRB_1991}the
phonon-bottleneck effect---because energy cannot be conserved in the
inter-shell relaxation process.
Finally, hole states do not form shells, with exception of flat
dots\cite{narvaez_JAP_2005} (height of about $20\;$\AA), and the splitting
between hole states is about $1$-$20\;{\rm meV}$, thus comparable to acoustic
phonon frequencies. Given these general characteristics, the main electron-
and hole-relaxation channels within the dot are:

(a) {\em Hole thermalization.} The hole relaxes to $h_0$, most likely via
electron--acoustic-phonon emission.  Such a hole relaxation has been found to
occur within sub-${\rm ps}$ times.\cite{sosnowski_PRB_1998,urayama_PRL_2001}
Moreover, Quochi and co-workers showed that the hole relaxation time depends
strongly on temperature: $20\;{\rm ps}$ at $60\;{\rm K}$ and $0.8\;{\rm ps}$
at $300\;{\rm K}$ (Ref. \onlinecite{quochi_PhysicaB_2002}).
Note that in CdSe colloidal dots the existence of energy gaps of $\sim
60\;{\rm meV}$ {\em within} the valence-band quasi-continuum was shown
experimentally\cite{xu_PRB_2002} and theoretically\cite{califano_NanoL_2003}
to slow down the hole thermalization.

(b) {\em Intra-shell electron relaxation} (e.g. $P_2\rightarrow P_1$; Fig.
\ref{Fig_1}). The electron relaxes from $P_2\rightarrow P_1$ ($1$-$6\;{\rm
  meV}$ splitting), or between magnetic-field split states, via acoustic
phonon emission. From optical pumb-probe measurements, Zibik {\em et al.} have
recently deduced relaxation times of $15\;{\rm ps}$ and $35\;{\rm ps}$ for
$P_1$-$P_2$ splittings of $3.7\;{\rm meV}$ and $5.5\;{\rm meV}$ (Ref.
\onlinecite{zibik_PhysicaE_2005a}), respectively. A model calculation that
adopts longitudinal acoustic phonon emission predicts, correspondingly, values
of $8\;{\rm ps}$ and $34\;{\rm ps}$.\cite{zibik_PhysicaE_2005a} 

(c) {\em Inter-shell electron relaxation for sole carrier and for
  electron-hole pair} (e.g. $P \rightarrow S$; Fig.  \ref{Fig_1}) within the
$40$-$60\;{\rm meV}$ separating the electronic shells.  This relaxation is
different if an electron-hole pair is present or just a sole electron (doped
dot).
As expected from the phonon bottleneck effect, inter-shell relaxation in
(In,Ga)As/GaAs dots has been observed to be slow by Urayama and
co-workers\cite{urayama_PRL_2001} (relaxation time of $\sim 750\;{\rm
  ps}$) as well as Heitz and co-workers\cite{heitz_PRB_2001} ($7.7\;{\rm ns}$).
In contrast, time-resolved optical measurements have clearly demonstrated that
this inter-shell decay is a {\em fast} process whether a hole is present or
not.  For instance, in experiments in which both an electron and hole are
present,
M\"uller {\em et al.} have found decay times of $4.7\;{\rm ps}$ at $5\;{\rm K}$
and $2.8$-$1.5\;{\rm ps}$ (depending upon excitation power) at
room-temperature in interband-pump--intraband-probe experiments (Ref.
\onlinecite{muller_APL_2003});
Boogart {\em et al.} found $19\;{\rm ps}$ (low intensity) and $9\;{\rm ps}$
(high intensity) within $5\;{\rm K}$ and $77\;{\rm K}$, but $7\;{\rm ps}$
(high intensity) at room-temperature, in time-resolved pump-probe differential
reflectance spectroscopy (Ref. \onlinecite{bogaart_icps27_2005});
Sosnowski {\em et al.} found $5\;{\rm ps}$ at $10\;{\rm K}$ in pump-probe
differential  transmission experiments (Ref. \onlinecite{sosnowski_PRB_1998});
De Giorgi {\em et al.} found $6.5\;{\rm ps}$ at $4\;{\rm K}$ ($3.0\;{\rm ps}$
at high intensity) and $3.5\;{\rm ps}$ at room-temperature in time-resolved PL
upconversion experiments (Ref. \onlinecite{de_giorgi_APL_2001});
with the same experimental technique, applied to large ($b=350\;${\AA},
$h=110\;${\AA}) and small ($b=250\;${\AA}, $h=30\;${\AA}) dots, Boggess {\em
  et al} found, respectively, $1\;{\rm ps}$ and $7\;{\rm ps}$ below $100\;{\rm
  K}$, and $\sim 2.5\;{\rm ps}$ at $200\;{\rm K}$ and $6\;{\rm ps}$ at
$150\;{\rm K}$ (Ref. \onlinecite{boggess_APL_2001});
while Siegert {\em et al.} found that at $80\;{\rm K}$ the $D\rightarrow S$
decay time corresponds to $7\;{\rm ps}$, $3\;{\rm ps}$, and $2\;{\rm ps}$ for
undoped, n-doped, and p-doped dots, respectively (Ref.
\onlinecite{siegert_PRB_2005}).
On the other hand, when a sole electron is present and no hole, the
inter-shell relaxation time slows down by a factor of about 2-10.
For instance, in $n$-doped (In,Ga)As/GaAs quantum dots the low-temperature
$P\rightarrow S$ relaxation time has been extracted from pump-probe infra-red
spectroscopy and is in the range of $20$-$65\;{\rm ps}$ in the experiments of
Zibik {\em et al.} (Ref.  \onlinecite{zibik_PRB_2004}) and $40$-$70\;{\rm ps}$
in the experiments of Sauvage {\em et al.} (Ref.
\onlinecite{sauvage_PRL_2002}). In the latter, the room-temperature
$P\rightarrow S$ relaxation is $37\;{\rm ps}$ for $\Delta(S-P)\simeq
54.5\;{\rm meV}$. Note that in earlier pump-probe interband absorption
experiments at high excitation Sauvage {\em et al.} found a relaxation time of
$3\;{\rm ps}$ at room temperature (Ref.  \onlinecite{sauvage_APL_1998}).
The situation is similar in colloidal dots such as CdSe, where the
$P\rightarrow S$ inter-shell relaxation in the absence of a hole slows down to
$\sim 10\;{\rm ps}$ (Ref. \onlinecite{guyot-sionnest_JCP_2005}), relative to
$\sim 1\;{\rm ps}$ when an electron-hole pair is present.

Several relaxation mechanisms have been proposed as responsible for the fast
inter-shell relaxation: multi-phonon emission,\cite{inoshita_PRB_1992} Auger
(carrier-carrier)
scattering,\cite{bockelmann_PRB_1992,jiang_PhysicaE_1998,ferreira_APL_1999,nielsen_PRB_2004}
and polaron
relaxation\cite{inoshita_PRB_1997,verzelen_PRB_2000,jacak_PRB_2002,seebeck_PRB_2005}.
(We discuss the Auger and polaron models in Sec. \ref{Auger+polaron}.)

\subsection{Thermal escape of carriers from dot}

Upon increasing temperature, the photoexcited electron and hole escape the
confined states of the dot [Fig. \ref{Fig_1}(d)].  Thermal depopulation has
been found to be significant at temperatures $T>100\;{\rm
  K}$.\cite{de_giorgi_APL_2001,urayama_APL_2002,norris_JPD_2005} However,
Heitz and co-workers have found the onset to be $200\;{\rm
  K}$.\cite{heitz_PhysicaB_1999} In {\em n}-doped InAs/GaAs dots, Bras and
co-workers showed that thermal depopulation becomes significant above
$70\;{\rm K}$ (Ref. \onlinecite{bras_APL_2002}).

\section{Auger and polaron mechanisms for $P\rightarrow S$ inter-shell decay}
\label{Auger+polaron}

\subsection{Auger relaxation via electron-hole scattering}

Figure \ref{Fig_1}(c) illustrates this process whereby the hot electron decays
by scattering a low-lying photoexcited hole into deep hole states like $h_k$.
Scattering takes place via the electron-hole Coulomb interaction, so this
relaxation process does not take place in the absence of a photexcited hole.
For the mechanism to be effective it requires energy conservation: The excess
energy of the electron has to be elastically transfered to the hole [as
sketched in Fig. \ref{Fig_1}, where ${\cal E}^{\,(e)}_1-{\cal
  E}^{\,(e)}_0={\cal E}^{\,(h)}_0-{\cal E}^{\,(h)}_k$].  On the other hand,
electronic level broadening due to phonons effectively relaxes this stringent
condition.\cite{sosnowski_PRB_1998}
In (In,Ga)As/GaAs self-assembled quantum dots the ${\cal E}^{(e)}_{P}-{\cal
  E}^{(e)}_S \sim 50\;{\rm meV}$ whereas in CdSe colloidal dots ${\cal
  E}^{(e)}_{P}-{\cal E}^{(e)}_S \sim 300\;{\rm meV}$. In the latter case the
$P\rightarrow S$ decay via Auger process is highly
effective.\cite{wang_PRL_2003,klimov_JPCB_2000,klimov_PRL_1998,guyot-sionnest_PRB_1999}
In fact, Hendry et al. \cite{hendry_PRL_2006} have demonstrated the validity
of the electron-hole Auger mechanism for $P\rightarrow S$ relaxation in CdSe
dots by measuring directly the hole thermalization time (Sec. II C) versus the
electron excess energy.  Moreover, in Ref. 48 Guyot-Sionnest and co-workers
have shown that in CdSe dots the $P\rightarrow S$ relaxation of electrons is
slowed down upon inducing hole trapping at the surface of the dots. This is
strong evidence in favor of relaxation due to electron-hole Auger scattering.
The effectiveness of the Auger mechanism for $P\rightarrow S$ relaxation in
self-assembled dots has been previously addressed within model Hamiltonians
only.\cite{jiang_PhysicaE_1998,ferreira_APL_1999} Here it will be calculated
by using a fully atomistic approach.
When the hole is absent (due to its capture by a hole-quencher, or when only
an electron is injected into the dot) the Auger mechanism is not possible. 
In CdSe colloidal dots the alternative mechanism corresponds to the coupling
of the electrons in the dot with virtual phonons of the
environment.\cite{guyot-sionnest_JCP_2005}. In (In,Ga)As/GaAs self-assembled
dots the polaron decay has been proposed
instead.\cite{inoshita_PRB_1997,verzelen_PRB_2000,seebeck_PRB_2005}

\subsection{Polaron decay for a single excited electron (no hole)}

This mechanism has been invoked to explain the electron relaxation to state
$e_0$ in the {\em absence} of a hole.
The confined electron states are assumed to be strongly coupled with the
continuum of states arising from the phonon replicas of the localized states
(e.g. $S$, $P$), thereby, forming stable polaron states. In turn, these
polaron states relax when the phonon component of the polaron relaxes due to
phonon anharmonicity.\cite{verzelen_PRB_2000} Thus, assuming that the phonon
component of the polaron originates from LO phonons, the phonon-bottleneck is
circumvented by the emission of an LO and a TA phonon. 
This mechanism requires that the $P$-$S$ energy difference be of the order of
the zone-center optical phonon energy. In colloidal dots ${\cal E}(P)-{\cal
  E}(S)\sim 200$-$500\;{\rm meV}$ for electrons while $\hbar\omega_{LO}\sim
30\;{\rm meV}$, so the polaron decay mechanism is not possible. On the other
hand, for holes in colloidal dots ${\cal E}(P)-{\cal E}(S)\sim 10$-$30\;{\rm
  meV}$, which would make the polaron decay possible. In (In,Ga)As/GaAs
self-assembled dots, ${\cal E}(P)-{\cal E}(S)\sim 50\;{\rm meV}$ for electrons
and ranges from $5$-$20\;{\rm meV}$ for holes while $\hbar\omega_{LO}\sim
30\;{\rm meV}$, thus making the polaron decay feasible.

In the case of the inter-shell $P\rightarrow S$ transition in (In,Ga)As/GaAs, the
polaron state has been predicted to relax within a few
picoseconds,\cite{verzelen_PRB_2000} leaving the excited electron in the $S$
state.
This model explains the observed relaxation times in the absence of a hole
(Sec. \ref{inter-shell_decay}).\cite{note-100}
Further data that has been taken as evidence of the polaron model in
(In,Ga)As/GaAs dots corresponds to the anticrossings in the energies of
allowed magneto-photoluminescence transitions as the field is
swept.\cite{hameau_PRL_1999} The magnitude of the anticrossings ($\sim 3\;{\rm
  meV}$) present in the spectra is consistent with those predicted by the
polaron model (Ref. \onlinecite{hameau_PRL_1999}).
We note that in low-symmetry dots all states have the same $a_1$-symmetry even
without phonon displacements, and therefore they would anticross in the
presence of a magnetic field. Whether the reason for lowering the symmetry to
$a_1$ is phonon coupling or simply the correct atomistic dot symmetry of the
non-vibrating dot remains to be determined.

\section{Calculation of Auger cooling due to  electron-hole scattering}
\label{Auger_calculation}

We have calculated the Auger cooling lifetime of electrons in In$_{\rm
  0.6}$Ga$_{\rm 0.4}$As/GaAs quantum dots within a pseudopotential-based
atomistic approach\cite{zunger_pssb_2001} in order to establish if this
mechanism leads to $P\rightarrow S$ decay times within magnitude needed to
explain low-excitation experiments in which a photoexcited hole is present.

\subsection{Method of calculation}
\label{method}

%
%
We begin by calculating the single-particle ladder $\{e_0,\,e_1,\,e_2,\dots\}$
and $\{h_0,\,h_1,\,h_2,\dots\}$ of electron and hole states, respectively, of
the (In,Ga)As/GaAs quantum dot. The wave function $\psi_j$ and energy ${\cal
  E}_j$ of these states are solutions of the atomistic single-particle
Schr\"odinger equation

\begin{equation}
\label{SP.Shrodinger}
\{-\frac{1}{2}\nabla^2+V_{SO}+\sum_{l,\alpha}\,v_{\alpha}({\bf R}-{\bf R}_{l,\alpha})\}\psi_j={\cal
  E}_j\,\psi_j.
\end{equation}

\noindent Here, the actual potential of the solid (dot+GaAs barrier) is 
described by a superposition of (semiempirical) screened pseudopotentials
$v_{\alpha}$ for atom of type $\alpha$ (In,Ga,As) with position ${\bf
  R}_{l,\alpha}$ within the dot or barrier, and a non-local pseudopotential
$V_{SO}$ that accounts for the spin-orbit interaction.\cite{williamson_PRB_2001} 
To solve Eq. (\ref{SP.Shrodinger}), we expand $\psi_j$ in a linear combination
of Bloch bands $u^{(M)}_{n,{\bf k}}({\bf R})$ of material $M$ (InAs, GaAs),
with wave vector {\bf k} and band index $n$, subjected to strain
$\tilde\varepsilon$:\cite{wang_PRB_1999}

\begin{equation}
\psi_j({\bf R})=\sum_{M}\sum_{n,{\bf k}}\,C^{(j)}_{n,{\bf
    k};M}\,u^{(M)}_{n,{\bf k};\varepsilon}({\bf R}) .
\end{equation}

\noindent This expansion has a main advantage over a plane-wave expansion: 
The Bloch bands $u^{(M)}_{n,{\bf k};\varepsilon}({\bf R})$ can be intuitively
chosen, which reduces the computational demand
significantly.\cite{wang_PRB_1999}
%
%
%
To calculate the electron Auger cooling lifetime $\tau(P\rightarrow S)$
due to electron-hole scattering at low temperatures, we proceed in two
steps. 

\subsubsection{Calculation of the Auger scattering rates for individulal
  electron-hole configurations}

We consider as initial electron-hole configurations
$|e_{i}h_j\rangle$ those corresponding to the electron in the $P$-shell
states $\{e_1,e_2\}$ and the hole in low-lying states $h_j$; and as the
final scattering states those that correspond to an electron occupying the
$S$-shell state $e_0$ and a hole in a deep state $h_k$ [Fig. \ref{Fig_1}(c)],
i.e $|e_0h_k\rangle$.
Then, we calculate the net, characteristic Auger scattering rate of the
transition $|e_{i}\rangle\rightarrow|e_0\rangle$ ($i=1,2$), with a hole in
state $h_{j}$, by using Fermi's golden rule:

\begin{equation}
\label{Eq_1}
\frac{1}{\tau_{i_h}(e_{i}\rightarrow e_0)}=\frac{2\pi}{\hbar}\sum_{k}\,|J^{\,(eh)}_{ij;0k}|^2\,\delta[E(i;j)-E(0;k)].
\end{equation}

\noindent Here, $E(i_e;i_h)$ and $E(0;k)$ correspond to
the many-particle energy of the initial and final state, respectively,
calculated at the single-configuration level of approximation.\cite{note_02}
The electron-hole Coulomb scattering matrix elements $J^{\,(eh)}_{i_ei_h;0k}$
are given by

\begin{equation}
\label{Coulomb_integrals}
J^{\,(eh)}_{ij;0k}=\int\int\,d{\bf R}d{\bf R}'\,
\frac{[\psi^{(h)}_{j}({\bf R})]^{*}[\psi^{(e)}_{i}({\bf
    R}')]^{*}\psi^{(e)}_{0}({\bf R}')\psi^{(h)}_{k}({\bf R})}
{\epsilon({\bf R},{\bf R}')|{\bf R}-{\bf R}'|},
\end{equation}

\noindent where $\epsilon({\bf R},{\bf R}')$ is the microscopic dielectric
function derived by Resta.\cite{resta_PRB_1977} Note that in the actual
computations, we introduce a phenomenological broadening $\Gamma$ of the final
states that allow us to replace $\delta(x)$ in Eq. (\ref{Eq_1}) with a
Gaussian function $(\Gamma\sqrt{2\pi})^{-1}\,\exp(-x^2/2\Gamma\,^2)$. One
should understand $\Gamma$ as a phenomenological way to account for the
phonon-induced (e.g. phonon broadening) finite lifetime $\tau_h$ of the
excited single-particle hole states: $\Gamma\sim 2\pi\hbar/\tau_h$.
Considering that experimentally the relaxation of a hole in the wetting layer
to $h_0$ takes about $0.6{\rm ps}$ (Ref.  \onlinecite{sosnowski_PRB_1998}), we
estimate a lower bound for $\Gamma$ of $10\;{\rm meV}$. The phenomenological
parameter $\Gamma$ has been used in previous calculations (Refs.
\onlinecite{jiang_PhysicaE_1998} and \onlinecite{wang_PRL_2003}).

\begin{figure}
  \includegraphics[width=8.5cm]{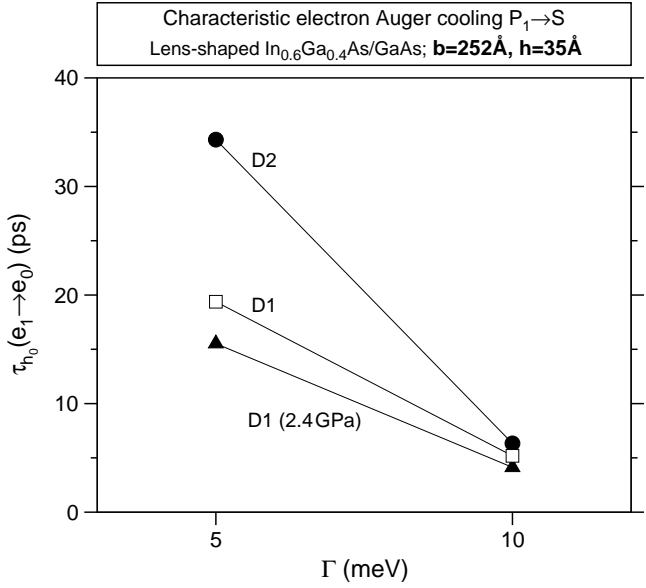}
\caption{{\label{Fig_2}}Electronic Auger cooling characteristic lifetime
  $\tau_{h_0}(e_1\rightarrow e_0)$
  calculated with two phenomenological broadenings---$\Gamma=5\;{\rm meV}$ and
  $10\;{\rm meV}$---for dots of same size (b,h)=($252\;$\AA,$35\;$\AA). Dots
  D1 (open squares) and D2 (solid circles) correspond to different random
  alloy disorder realizations.}
\end{figure}

%
%
Figure \ref{Fig_2} shows the characteristic Auger relaxation lifetime
$\tau_{h_0}(e_1\rightarrow e_0)$ calculated for two values of $\Gamma$ in two
lens-shaped In$_{\rm 0.6}$Ga$_{\rm 0.4}$As/GaAs quantum dots---D1 and D2---of
size ($252\;$\AA, $35\;$\AA). These dots differ only in the random alloy
disorder realization.
For a phenomenological broadening $\Gamma=5\,{\rm meV}$,
$\tau_{D1}(P\rightarrow S) \sim 20\;{\rm ps}$ and $\tau_{D_2}(P\rightarrow
S)\sim 35\;{\rm ps}$. The strong difference shows that
$\tau_{h_0}(e_1\rightarrow e_0)$ depends strongly upon the energy structure of
the final states. For a more plausible value of the broadening,
$\Gamma=10\;{\rm meV}$, $\tau_{h_0}(e_1\rightarrow e_0)\sim 5\;{\rm ps}$ for
both dots.\cite{note-200} In addition, we find that $\tau_{h_0}(e_1\rightarrow e_0) \simeq
\tau_{h_0}(e_2\rightarrow e_0)$; D2 presents a difference of $1.5\;{\rm ps}$
among these lifetimes.
We also show, for a comparison, $\tau_{h_0}(e_1\rightarrow e_0)$ for dot D1
under a hydrostatic pressure of $2.4\;{\rm GPa}$. Because this pressure does
not change significantly the intraband energy structure of the confined
states, but it primarily increases the localization of their wave
functions,\cite{narvaez_PRB_2005a} the characteristic relaxation lifetime is
smaller than at ambient pressure.

%
%

\subsubsection{Solution of the rate equations describing the $P\rightarrow
  S$ electron relaxation}

Once we have calculated the characteristic times
$\tau_{i_h}(e_{i_e}\rightarrow e_0)$, we notice that (i) at low temperatures
($k_BT \ll {\cal E}^{(h)}_{1}-{\cal E}^{(h)}_{0}$) there are two relevant
initial electron-hole configurations $|1\rangle=|e_1h_0\rangle$ and
$|2\rangle=|e_2h_0\rangle$ that decay to a single scattering configuration
$|s\rangle=|e_0h_k\rangle$. (ii) In addition, due to the $P\rightarrow P$
intra-shell relaxation, configuration $|2\rangle$ decays to $|1\rangle$ with a
relaxation time $\tau(e_2\rightarrow e_1)=\tau_{21}$ between $15\,{\rm ps}$
and $35\,{\rm ps}$.\cite{zibik_PRB_2004}
Thus, we find the time-dependent occupation of $n_1$, $n_2$, and $n_S$ by
solving numerically the following set of rate equations.

\begin{eqnarray}
\frac{d n_1}{d t}&=& -[\gamma^{(+)}+(\tau_{h_0}(e_1\rightarrow e_0))^{-1}]n_1+\gamma^{(-)}n_2 \nonumber\\
\frac{d n_2}{d t}&=& -[\gamma^{(-)}+[\tau_{h_0}(e_2\rightarrow e_0)]^{-1}]n_2+\gamma^{(+)}n_1 \nonumber\\
\frac{d n_s}{d t}&=&[\tau_{h_0}(e_1\rightarrow e_0)]^{-1}n_1+[\tau_{h_0}(e_2\rightarrow e_0)]^{-1}n_2
\end{eqnarray}

\noindent with initial conditions taken to be $n_1(0)=n_2(0)=1/2$ and
$n_S(0)=0$. These initial conditions reflect the fact that the electrons
captured in the dot have the same probability to decay to $P_1$ or $P_2$ (see
Sec. \ref{inter-shell_decay}).

Here, $\gamma^{(+)}$ and $\gamma^{(-)}$ are the rates of transitions
$n_1\rightarrow n_2$ and $n_2\rightarrow n_1$, respectively, with

\begin{equation}
\gamma^{(+)}=\frac{1}{\tau_{21}}[\exp(\Delta E/k_B T)-1]^{-1}
\end{equation}

\noindent and 

\begin{equation}
\gamma^{(-)}=\frac{1}{\tau_{21}}[1+(\exp(\Delta E/k_B T)-1)^{-1}];
\end{equation}

\noindent where $\Delta E=E(2;0)-E(1;0)$.
Finally, we extract electron Auger relaxation $\tau(P\rightarrow S)$ by
fitting the time-dependence of the occupation probability $n_s$ to the
expression $1-\exp[-t/\tau(P\rightarrow S)]$. For the characteristic times
$\tau_{h_0}(e_1\rightarrow e_0)$ and $\tau_{h_0}(e_2\rightarrow e_0)$
calculated with $\Gamma=10\;{\rm meV}$, and $\tau_{21}=15\;{\rm ps}$, the fit
is excellent.

%
%
\begin{figure}
  \includegraphics[width=8.5cm]{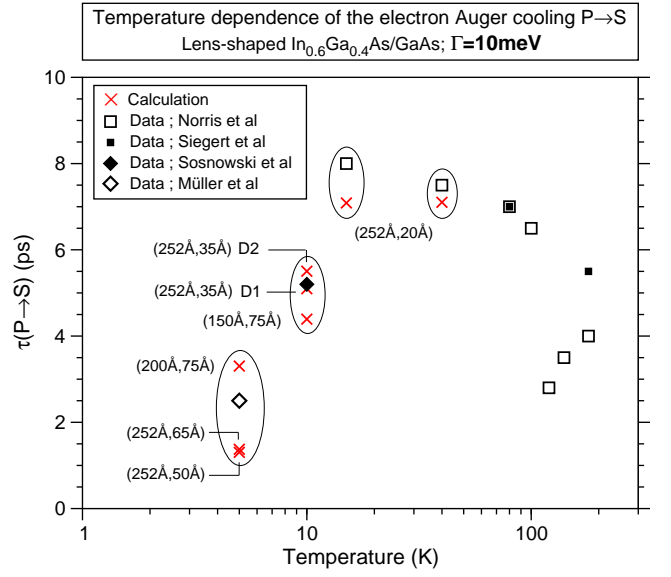}
\caption{{\label{Fig_3}}(Color online.) Auger cooling lifetime
  $\tau(P\rightarrow S)$ {\em vs}
  temperature for seven lens-shaped quantum dots of different sizes. The pair
  (b,h) indicates the base diameter and height of the dots. Data from
  Refs.
  [\onlinecite{norris_JPD_2005,siegert_PRB_2005,muller_APL_2003,sosnowski_PRB_1998}]
  are also shown.}
\end{figure}

%
%
\subsection{Predicted $\tau(P\rightarrow S)$ and comparison with data}

Figure \ref{Fig_3} shows $\tau(P\rightarrow S)$ versus temperature for
lens-shaped dots of different sizes [{\rm (base,$\,$height)}].  In these
calculations the broadening $\Gamma=10\,{\rm meV}$ is larger than the average
energy spacing of the relevant final states and $\tau_{21}=15\;{\rm ps}$. Two
features are prominent.
(i) $\tau(P\rightarrow S)$ decreases with both increasing height at a fixed
base and increasing base at a fixed height.
(ii) The Auger cooling lifetime of ($150\;$\AA,$75\;$\AA) is similar to that
of dots with size ($252\;$\AA,$35\;$\AA) due to their similar
single-configuration exciton gap (see below). {\em Comparison with data:} In
Fig. \ref{Fig_3} we also show data extracted from differential transmission
spectroscopy experiments (Ref.  \onlinecite{norris_JPD_2005}) and
time-resolved photoluminescence experiments (Refs.
[\onlinecite{siegert_PRB_2005,muller_APL_2003,sosnowski_PRB_1998}]) in
(In,Ga)As/GaAs dots appear as squares and diamonds. A comparison with our
calculated values shows the following.
(i) We find excellent agreement between our calculated $\tau(P\rightarrow S)$
for the ($252\;$\AA,$35\;$\AA) dot D1 and the value of $5.2\;{\rm ps}$ found
by Sosnowski and co-workers in differential transmission spectroscopy in
(In,Ga)As/GaAs dots with gap of $1.265\;{\rm eV}$ (Ref.
\onlinecite{sosnowski_PRB_1998}). Dot D2 and the dot with size ($150\;$\AA,
$75\;$\AA) also compare well with experiment.
(ii) The value of $2.5\;{\rm ps}$ for $\tau(P\rightarrow S)$ at $5\;{\rm K}$ (Fig.
\ref{Fig_3}) in InAs/GaAs dots with energy gap of $1.08\;{\rm eV}$ that has
been derived by M\"uller {\em et al.} (Ref.  \onlinecite{muller_APL_2003}) from
pump-probe intraband spectroscopy is in satisfactory agreement with our
predicted values for ($252\;$\AA, $50\;$\AA), ($252\;$\AA, $65\;$\AA), and
($200\;$\AA, $75\;$\AA) dots.
(iii) Our results for the flat dot ($h=20\;${\AA} and $35\;${\AA}) compare well
with the $\tau(P\rightarrow S)$ data of Norris {\em et al.} (Ref.
\onlinecite{norris_JPD_2005}) at low temperatures. The data of Siegert {\em et
  al} (Ref. \onlinecite{siegert_PRB_2005}) below $100\;{\rm K}$ is comparable
to our low-temperature predicted values. Note that Norris {\em et al.} have
found that above $100\;{\rm K}$ thermal escape of carriers [Fig.
\ref{Fig_1}(e)] is important, which explains the large abrupt reduction of the
Auger decay time seen in the data.\cite{norris_JPD_2005}

%
%
\begin{figure}
  \includegraphics[width=8.5cm]{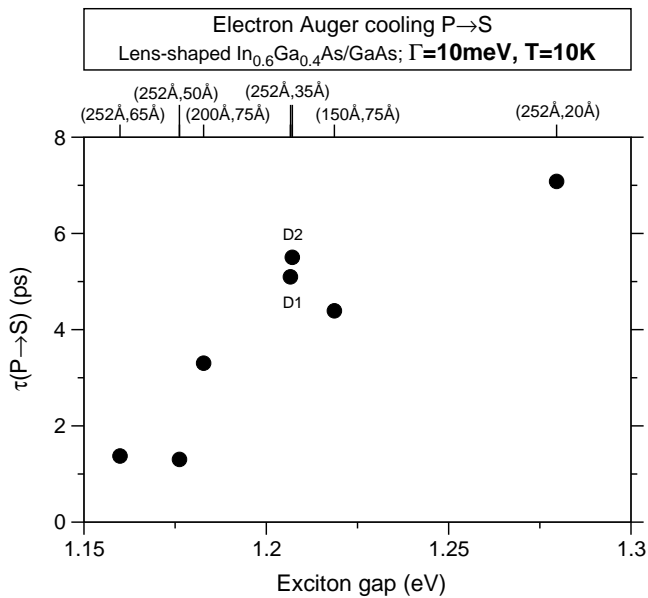}
\caption{{\label{Fig_4}}Calculated Auger-cooling
  lifetime $\tau(P\rightarrow S)$ at $T=10\;{\rm K}$ versus the
  single-configuration exciton gap for several lens-shaped quantum dots.}
\end{figure}

\subsection{Trend of $\tau(P\rightarrow S)$ with exciton gap}

Figure \ref{Fig_4}(a) shows the calculated low-temperature ($10\,{\rm K}$)
Auger relaxation lifetime as a function of the dot exciton gap for several
In$_{0.6}$Ga$_{0.4}$As/GaAs quantum dots.\cite{note_01}
Two important features emerge: (i) We find that $\tau(P\rightarrow S)$ ranges
from $1$-$7\;{\rm ps}$ and decreases with the gap of the dots.  As the $S$-$P$
splittings of the lens-shaped dots is nearly the same, we attribute the
reduction of $\tau(P\rightarrow S)$ to the {\em increase} of the joint density
of states

\begin{equation}
\label{JDOS}
g[E(i_e,i_h)]=\sum_{k}\delta[E(i_e;i_h)-E(0;k)]
\end{equation}

\noindent that takes place as the gap of the dot {\em decreases}, due to the
increase in the density of single-particle hole states.

\subsection{Comparison with other calculations for (In,Ga)As/GaAs dots}

We have compared our results with two {\em model} calculations. (i) The 8-band
${\bf k}\cdot {\bf p}$ calculation of Jiang and Singh (Ref.
\onlinecite{jiang_PhysicaE_1998}) and (ii) the parabolic, single-band
effective-mass calculation of Ferreira and Bastard (Ref.
\onlinecite{ferreira_APL_1999}).
Our results agree well with the calculation in (i). Namely, Jiang and Singh
show an increase of the characteristic Auger cooling lifetime with decreasing
$\Gamma$.
In addition, the results of Jiang and Sing compare satisfactorily (within a
factor of two) with the value of $\tau(P\rightarrow S)$ observed by Sosnowsky
{\em et al}.\cite{sosnowski_PRB_1998}
A direct comparison with (ii) is not fully applicable since Ferreira and
Bastard consider different initial states than those considered here (Sec.
\ref{method}). In particular, the starting electron-hole pair states
correspond, in our language, to $|e_1h_1\rangle$ and $|e_1h_2\rangle$.
However, it is interesting to see that Ferreira and Bastard find that the
Auger-cooling lifetime is within $0.1$ and $6\;{\rm ps}$. Moreover, depending
on the choice of initial e-h states, this lifetime either increases as gap decreases
(in contrast to our predictions; Fig. \ref{Fig_4}) or viceversa.

\subsection{Digression: Comparison with calculations and data for CdSe colloidal dots}

%
%
Wang {\em et al.}\cite{wang_PRL_2003} have calculated $\tau(P\rightarrow S)$
for CdSe colloidal dots using the same methodology as in this
paper---pseudopotential-based atomistic approach---finding, respectively,
relaxation times of $0.6\;{\rm ps}$ and $0.2\;{\rm ps}$ for a dots with radii
of $29\;${\AA} and $38\;${\AA}. These results show that in contrast to
In$_{\rm 0.6}$Ga$_{\rm 0.4}$As/GaAs dots, $\tau(P\rightarrow S)$ increases
with decreasing the dot gap. Moreover, for In$_{0.6}$Ga$_{0.4}$As/GaAs dots,
we predict $\tau(P\rightarrow S)$s that are about a factor of 10 slower.
The ${\bf k}\cdot{\bf p}$-based calculation of Efros and
co-workers\cite{efros_SSC_1995} predicts Auger decay lifetimes in CdSe
colloidal dots of $\sim 2\;{\rm ps}$ almost independently of dot size for
radii between $20\;${\AA} and $40\;${\AA}. While the magnitude of
$\tau(P\rightarrow S)$s that we find in InGaAs/GaAs is comparable to that of
Efros and co-workers, the gap dependence is strikingly different.
%
%
On the other hand, bleaching experiments in CdSe colloidal quantum dots show
that the Auger cooling lifetime of electrons is below a picosecond and {\em
  decreases} as the exciton gap {\em increases}.\cite{klimov_JPCB_2000} [Note
that the calculations of Wang and co-workers\cite{wnag_PRL_2003} capture reproduce these
experimental findings.]
We predict that $\tau(P\rightarrow S)\sim 1-7\;{\rm ps}$ in (In,Ga)As/GaAs
self-assembled quantum dots and shows the opposite gap dependence [Fig.
\ref{Fig_4}]. The gap dependence of $\tau(P\rightarrow S)$ in both colloidal
and self-assembled dots is dictated by the gap (size) dependence of (i) the
joint density of states [Eq. (\ref{JDOS})] and (ii) the magnitude of the
Coulomb scattering integrals [Eq. (\ref{Coulomb_integrals})].  While in
(In,Ga)As/GaAs self-assembled dots the changes with size in the joint density
of states prevails, in CdSe colloidal dots the changes of the Coulomb
integrals dictates the gap dependence of $\tau(P\rightarrow S)$.

\section{Summary}

We have discussed several dynamical processes that photoexcited electrons and
holes undergo in (In,Ga)As/GaAs self-assembled quantum dots, and calculated
the inter-shell $P$-to-$S$ electron decay lifetime in (In,Ga)As/GaAs
self-assembled dot due to Auger electron-hole scattering. When only an
electron (or only a hole) is present due to doping and this sole carrier is
excited by a photon, its decay must involve a non-Auger mechanism (perhaps
polaron decay). But when both an electron and hole are present we show that
this Auger cooling takes place within picoseconds, which makes it an efficient
inter-shell relaxation process compared to radiative recombination ($\sim {\rm
  ns}$). In addition, we predict that the lifetime $\tau (P\rightarrow S)$
increases with the exciton gap. Our pseudopotential-based calculations confirm
earlier predictions of simplified, {\em model} calculations. The values we
find for $\tau (P\rightarrow S)$ compare well with recent data in the presence
of photoexcited holes. This finding complemented with our review of the data
in the literature allows us to conclude that in the presence of a photoexcited
hole there is no need to invoke the alternative polaron-decay mechanism for
inter-shell electron relaxation. This conclusion could be tested in
(In,Ga)As/GaAs dots by measuring the rate of hole thermalization versus the
electron excess energy, or by measuring the electron relaxation rate after
modifying the surface of the dot so as to cause hole trapping. Finally, a
consistent picture of electron relaxation within quantum dots appears to
demand two relaxation mechanisms: electron-hole Auger scattering and polaron
decay.

\begin{acknowledgments}
  The authors thank Alberto Franceschetti (NREL) for useful discussions. This
  work was funded by U.S. DOE-SC-BES-DMS under Contract No.  DE-AC36-99GO10337
  to NREL.
\end{acknowledgments}


\begin{thebibliography}{100}
\bibitem{note_02} We disregard both the electron-hole exchange interaction,
  which has a magnitude of a few $\mu{\rm eV}$, and correlations that would
  arise from electron-hole configuration mixing.
\bibitem{note-200} Note that unless the microscopic character of the random
  alloy disorder changes significantly the small lifetime of the holes, the
  choice of $\Gamma$ is independent of disorder. We have made this assumption,
  as assessing the microscopic effects of alloy disorder on the characteristic
  Auger relaxation times lays beyond the scope of the present paper.
\bibitem{note_01} The dot exciton gap is calculated within the
  single-configuration approximation, which predicts a gap that is an
  excellent approximation to the actual fully-correlated exciton gap.
\end{thebibliography}
\end{document}